# Privacy as a Service in Digital Health

Xiang Su, Jarkko Hyysalo, Mika Rautiainen, Jukka Riekki, Jaakko Sauvola, Altti Ilari Maarala, and
Harri Honko

*Abstract*—Privacy is a key challenge for continued digitalization of health. The forthcoming European General Data Protection Regulation (GDPR) is transforming this challenge into regulatory directives. User consent provisioning and coordinating across data services will be the keys in addressing this challenge. We suggest a privacy-driven architecture that provides tools for providing user consent as a service. This enables managing and reusing private health information between a large amount of data sources, individuals and services, even when they are not known beforehand. The proposed architecture integrates data security and semantic descriptions into a trust query framework to provide the required interoperability and co-operation support for future health services. This approach provides benefits for all stakeholders through safer data management, cost and process savings, multi-provider services, and services based on emerging new business models.

## I. INTRODUCTION

DIGITALIZATION of health services and health data enables more efficient processes, cost savings, and creation of new services and business models, thus also improving the coverage, quality, safety and efficiency of health care. However, the multitude of different organizations providing and consuming data sources and services introduces challenges on the path towards the data-driven future of health care. One of the main challenges is privacy and specifically consent, that is, an indication of the data of the subject's wishes by which the subject signifies agreement to the processing of the subject's personal data, either by statement or by "clear affirmative action" [1].

Current mechanisms of informed consent in healthcare are still mostly static and paper-based [2], which does not comply with native digital health scenarios. The forthcoming European General Data Protection Regulation (GDPR) [1] reforms data protection rules and must be considered when services consuming personal data are developed. GDPR targets to protect and enable human-centric control of personal data; its main objective is to give citizens back the control over of their personal data and to simplify the regulatory environment for business utilization.

This paper suggests tackling these challenges by designing a system architecture, fit for the digitalized health services, that provides "privacy as a service", PRIAAS. Our work has been inspired by the MyData effort, depicted in a white paper [3]. Derived from this, our architecture offers a comprehensive solution approach for consent delivery and management in person-centric health data services infrastructure. The "privacy as a service" architecture requirements are inherited from extended MyData principles augmented with, e.g. GDPR and other health actor directives. We call these as the key requirements for human-centric processing and managing of personal information. In PRIAAS, the requirements are classified to five distinct categories: 1) CONTROL: Individuals shall have the right and practical means to manage their data and privacy according to GDPR attributes; 2) ACCESS: The data shall be easy to access and use; 3) TRANSLATION: There shall be a way to convert the data from single entities into a meaningful, machine readable resource that can be used to create new services; 4) INTEROPERABILITY: In support of open business environment, the shared data infrastructure shall enable the coordinated management of personal data, ensure interoperability, and provide easier means for different entities to comply with tightening data protection regulations; and 5) PROVISIONING: It shall allow individuals to change service providers and have a control over their data manager(s).

Our main contribution is a novel privacy as a service approach, with classification of essential requirements, followed by the practical PRIAAS system architecture for providing means to manage collected personal data according to the detailed directives set by GDPR and various health data case requirements. The proposed architecture enables users to have effortless control and ownership of their own data. The PRIAAS architecture is the first open GDPR conforming solution, developed and potentially widely utilized in an EU country, and endorsed by Finnish government spearhead agenda, for instance. In this article, we present human-centric principles, the system architecture, and a proof-of-concept about services consuming data from multiple sources.

## II. STANDARDS AND TECHNOLOGIES FOR CONSENT

Personal data has significant social, economic, and practical value. However, individuals currently have very limited control over their own data, often gathered by a large number of relevant actors. Although information is nowadays routinely shared digitally at the global level, mechanisms of consent remain traditional, static, vertical actor driven, and organized around national boundaries, specific service provider rules and legal frameworks [2]. These mechanisms are either paper-based signatures or online interactions, such as online forms, buttons, and opt-in checkboxes. They have clear weaknesses regarding scalability and interoperability, but mostly they don't comply with distributed health data service requirements.

These limitations motivate the development of novel data protection regulations, which return the control of an individual's personal data usage back to the data subject. The general goal of GDPR is to protect individuals against abuse of personal data. Personal data can be anything, like a name, a photo, an email address, bank details, posts on social networking websites, different types of static or non-static media or medical information. The basic ideas behind GDPR are that individuals control their own data and that trust is built in personal data services through a combination of transparency, interchangeability, public governance, respectable companies, public awareness, and secure technology. Control is realized through consents that determine what data services can fetch and how it can be processed.

Other relevant standards include User Managed Access (UMA) [5] and upcoming Minimum Viable Consent Record (MVCR) specification [6]. GDPR, UMA, and MVCR all share similar goals of giving individuals unified control points for authorizing who and what can get access to their digital data, content, and services. Moreover, they all consider simplicity, ease-of-use, user-centeredness, transparency, and standardization. While GDPR sets the legal framework that calls for explicit, unambiguous and informed consent, transparency and interoperability, UMA and MVCR offer technologies that can be used to address authorization and consent. They can also be used to address the requirements set by GDPR.

UMA is an OAuth 2.0 based access management protocol that enables individuals to have control over their personal data, content and services. UMA focuses on connecting a service providing an individual's personal data to a service consuming that data so that the individual can securely manage access to his/her data. We adopt several UMA protocol characteristics in our PRIAAS approach: 1) unified access control under a dedicated online service; 2) applying the same policies across multiple sites; 3) support for claims-based access policies, e.g. "over 18"; and 4) easy end-user management of access control. MVCR, in turn, specifies requirements for creating a legitimate digital consent record. MVCR aims at the minimal amount of information that individuals need to address for an explicit consent. Individuals can use the consent receipt to communicate with organizations about the consent details and the purposes it can be used to authorize data access.

Arnold et al. [7] conducted a literature review and found no single solution addressing all or even the majority of the issues of informed consent. However, we are aware of a recent implementation, e.g. ForgeRock Identity Platform [4], which addresses the evolving customer data privacy regulations based on UMA.

### III. Human-centric Data Management Principles

Europe is under a paradigm shift towards digital personal data management and processing. The current organization-centric data management model is being transformed to a model that allows for organizations to exchange data more flexibly, but on the other hand, gives the control of exchange and permissions to the hands of individuals. Since GDPR rules impose constraints over the design of human-centric systems, these rules can be considered as the foundation and requirements basis for the system architecture design.

Figure 1 illustrates this new paradigm from the perspective of an individual. The individual produces data to Services (S) that are designed to collect data (D) through a process imposed by an organizational entity, e.g. occupational healthcare that collects personal health records. Applications (blue belt) present interfaces to users that are increasingly involved in the organizational data collecting and utilization process. Aggregator Services (AS) are able to access different data repositories to add new value by correlating and analyzing data from different organizational sources (e.g. by combining health records from occupational and public healthcare).

To ensure trusted and fair utilization of data between organizations, GDPR imposes new user rights that enforce organizations to build more tools for control over the collected data. In the figure, numbers I-VIII represent GDPR imposed rights that can be considered as the most relevant in enabling a human-centric architecture in Europe: (I) the right for unambiguous consent; (II) the right that only relevant, necessary, accurate and legitimate data is processed in a specific, fair and transparent manner; (III) the right of access to one's own personal data; (IV) the right to be properly informed of personal data processing; (V) the right to rectification; (VI) the right to be protected against the use of personal data for automated profiling; (VII) the right to be forgotten; and (VIII) the right of security measures.

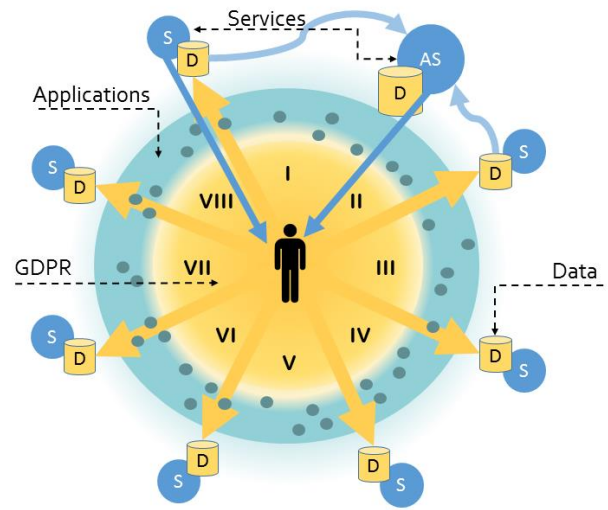

Fig. 1. The human-centric approach lets individuals control the use of their personal data over a variety of digital services.

The GDPR regulations create a foundation of requirements to our proposed architecture that enables informed user consent delivery and management for data exchange. This architecture is also at the core of the MyData approach [3], a novel procedural approach to describe personal data management that combines the digital rights of individuals with the needs of organizations and industry. At its core, MyData is a consent centric approach that benefits individuals, organizations, and society at large. MyData approach bridges the centrally controlled multi-organizational data silos and the fully decentralized systems currently existing on the Web.

In practice, MyData approach is based on a specific operating entity, called MyData Operator that allows users to arrange and manage data exchange between Sources and Sinks, the entities that take care of storing, representing and processing data for user applications. Portability and minimal service provider lock-in are emphasized, hence individuals can choose and migrate MyData Operators. MyData Operator is a novel, central entity for service registration and consent management that is compliant with the GDPR.

Figure 2 (left) illustrates the principles and roles in MyData approach. The flow of consents is separated from the data flow. Consents are managed through MyData Account but the data itself is not necessarily streamed through the server hosting this account. Data sources and services consuming data exchange information with the MyData Account using MyData compliant APIs. Individual end-users may grant access and give or cancel permissions for multiple data sources and services using this centralized interface. Any service provider can build a MyData API and enable their service to be connected with MyData Accounts. Moreover, MyData architecture aims at allowing individuals to switch operators in a convenient way (see, Figure 2 right). This account portability increases the trustworthiness of MyData approach.

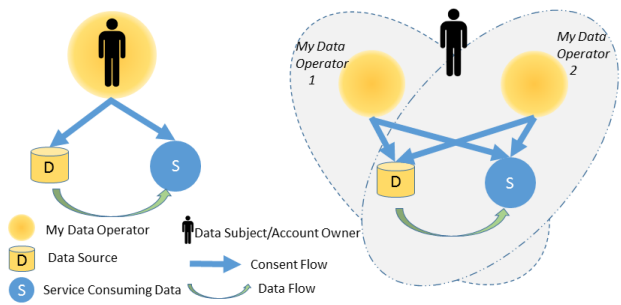

Fig. 2. Left: Key roles of MyData. Right: Account portability.

## IV. ARCHITECTURE FOR PROVIDING PRIVACY AS A SERVICE

One of the central ideas in GDPR is that the control resides on the individual, who must be well aware where his/her data will be used before granting consent for data utilization. This requires centralized consent management in a distributed environment. The proposed consent management solution can be understood as privacy as a service (PRIAAS), offering flexible ongoing monitoring and governance mechanisms instead of a one-off static consent. We are not aware of any other similar solution in the extent of the massively (Internet-wide) distributed private data environments; the current practice is to handle consent on a service-by-service, or vertical service operator provisioning basis.

The proposed consent management framework is based on roles with different responsibilities and liabilities. It is inspired by UMA specification, borrowing some concepts and naming conventions from UMA protocol (like Protection API, Authorization API and the concept of Resource set). However, our consent management framework does not conform to UMA protocol flow. Consents are managed by MyData Operator(s), a novel concept that lets users arrange and manage data exchange between Sources and Sinks, and which simplifies the authorization process compared to the UMA approach. MyData Account is another key element to enable individuals' to view, manage and control their consents easily through one operator's user interface in a transparent and standardized way. For an individual, MyData Account is a single hub for managing personal data across different organizations in a horizontal manner, and for providing tools for consent management and permissions of data use. Standardization of the process also facilitates interoperability and discovery.

User accounts can be held and managed by one or more trusted MyData Operators. MyData Operators provide logical paths to data owners (individuals) to control their personal data in complex environments of numerous data sources and consumers. The user accounts can be provided either by organizations acting on behalf of the individuals, or by the individuals, who can setup their own user account services (similarly as hosting a personal web service or e-mail server), offering the accounts as a service.

Figure 3 presents the proposed architecture for providing PRIAAS. This architecture is used at the core of MyData Operators. MyData Operators provide Web APIs that are used to register Sources and Sinks through Protection and Authorization APIs. Respectively, services that implement the roles of Sources and Sinks are required to provide APIs for exchanging consent information, while the MyData Operator acts as a broker. In practice, Sources can use these APIs to enquire the trust of Sinks before providing access to data. Actual data exchange happens between Sources and Sinks without the involvement of MyData Operator. This allows flexibility to the data architecture and keeps MyData Operator role lightweight. This way, a variety of organizations can establish and maintain an operator service; an important characteristic that helps to speed up the building of ecosystem around MyData approach.

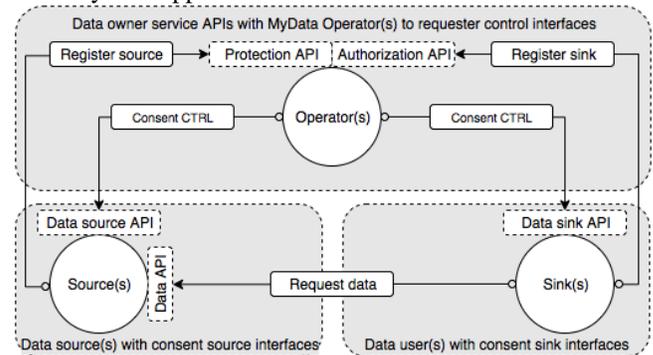

Fig. 3. Consent management architecture clarifying roles, responsibilities and liabilities.

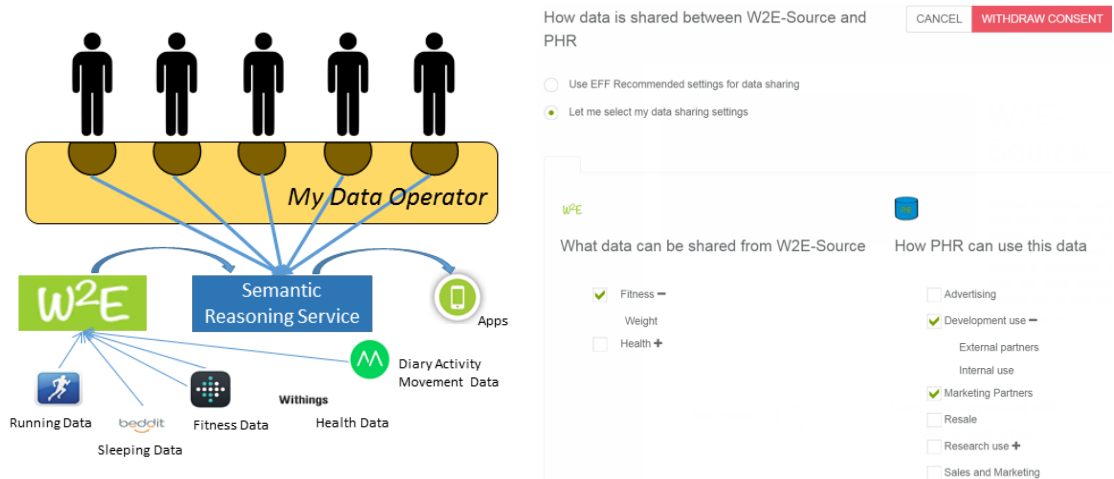

Fig. 4. Left: Privacy as a Service in Multiple provider service. Right: Consent interface.

Our open architecture enables developers to build access and services via public programming interfaces and libraries. The consent management architecture separates the flows of data based on the purpose and usage rights. Consent permissions, e.g. protection, authorization and control are managed via a separated set of interfaces and programming instances, allowing new consent operators, services and applications to evolve. Personal data flows between sources and sinks, with eligible operator permissions. Moreover, user account management is separated from consent services and data flows. Hence, the operator never stores any personal data generated by any source but only acts as a trusted consent manager and exposes the rights or limits to the utilization of an individual's data, on behalf of an individual.

Our authorization is based on a centralized authorization server similar to UMA. As resource servers and clients are always discoverable and trusted via their registration to the service registry, our authorization flow requires fewer messages compared to full UMA flow, as we no longer need to introduce the parties to each other in the beginning of authorization.

## V. Proof-of-Concept: Multi-Provider Services

To validate the feasibility of our consent management approach and architectural solution, we developed a proof-of-concept implementation, where users are able to authorize data sinks and sources to exchange data in a secure and trusted manner. Data is exchanged through common data APIs and consent is transmitted through OAuth protocol. The proof-of-concept is constructed into separate implementations for data operator(s), data source(s) and data sink(s), each with distinct and interoperable roles and responsibilities in terms of consent management regulation. Our solution enables the owner of a user account to complete necessary actions to establish a consented data flow of one's data from a source to a sink.

MyData compliant data sources provide data in a machine-readable format (JSON) through RESTful interfaces. The architecture also enables the data sources and data users to exchange information using the MyData Operator and its defined core transactions.

We implemented a health and wellness recommendation service based on multiple data sources. As shown in Figure 4 (left), Operator manages the consents and data authorizations of end users, by interacting with modified W2E [8] and Semantic Reasoning Service components. End users must have a user account at each of these services. W2E MyData proxy and Semantic Reasoning Service enable privacy preservation through pseudonymization and separation of consent and data flows. W2E acts as a proxy and accesses data from multiple non-MyData compliant data sources, mostly backend servers of various health and wellness device manufacturers, and delivers the data for the reasoning service based on consent of the end users – from the operator which manages the consent registry of each end user. A screenshot of consent management interface is presented in Figure 4 (right), where users can connect Data Sources and Data Sinks, and specify how their data can be utilized. In this case, a user wants to share his Fitness data from W2E with Personal Health Record (PHR). Hence, PHR can fetch this data from W2E and to utilize (analyze, refine, etc.) it in order to provide value for the user.

The reasoning service performs inference tasks based on ontologies and rules and suggests health and wellness recommendations to the end user applications through an API. These recommendations are concluded from a set of rules (Table 1) based on the current Finnish healthcare guidelines publicly available from the Finnish healthcare and medical databases. The rules infer a person's overall health, diabetes risk and stress level from data fetched from multiple data sources. This allows a layer of trust to applications: the semantic reasoning service can be maintained by officially authorized organizations that guarantee the validity of data driven reasoning service for third party applications interacting with end users.

TABLE I
A SELECTION OF RULES FOR INFERRING HEALTH RELATED CONDITIONS.

| Fact | Clause |
|---|---|
| TotalExercise | Exercise hasTimeStamp between(x,y) ∧ hasDuration ?d →TotalExercise hasDuration sum(?d) ∧ hasMeasurementDuration(y-x) |
| LowExerciseAmount | TotalExercise hasDuration ?d ∧ hasMeasurementDuration ?md ∧ ?d/?md < 0.04 → LowExerciseAmount |
| EnoughIntenseExercise | Exercise rdf:type IntenseExercise hasTimeStamp between(x,y) ∧ count>3 ∧ sum(hasDuration)/hasMeasurementDuration > 0.0074 → EnoughIntenseExercise |
| BMIIndex | (Weight/Height^2)*703 ?bmi → BMIIndex hasBMI ?bmi |
| Obesity | BodyMassIndex > 29.9 → Obesity |
| EfficientSleep | SleepEfficiency > 84 → EfficientSleep |
| OptimalBP | SystolicBloodPressure < 120 ∧ DiastolicBloodPressure < 80 → OptimalBP |
| HypertensionDeg1 | 159 > SystolicBloodPressure > 140 ∧ 99 > DiastolicBloodPressure > 90 → HypertensionDeg1 |
| DiagnosedHypertension | (HypertensionDeg1 ∨ HypertensionDeg2 ∨ HypertensionDeg3) hasTimestamp between(x,y) ∧ avg(hasSystolic) > 140 ∧ avg(hasDiastolic) > 90 → DiagnosedHypertension |
| UnhealthyDiet | Purchases hasTimestamp between(x,y) ∧ rdfs:subClassOf Fruits_Berries_Vegetables count+1 ∧ count < 2TimesPerWeek → UnhealthyDiet |
| VeryHighType2DiabetesRisk | Age>64 ∧ Obesity ∧ DiagnoseHighBP ∧ (NotEnoughIntenseExercise ∧ NotEnoughModerateExercise) ∧ FamilyMember hasDiagnosedDiabetes ∧ HighBloodGlucose ∧ UnhealthyDiet → VeryHighType2DiabetesRisk |
| OptimalHealth | Normalweight ∧ (EnoughIntenseExercise ∨ EnoughModerateExercise) ∧ NormalBP ∨ OptimalPB ∧ EfficientSleep → OptimalHealth |
| Stressed | HypertensionDeg1 ∨ HypertensionDeg2 ∧ InefficientSleep hasTimestamp between(x, y) → Stressed ∧ Relax |
| ReduceTraining | Underweight ∧ HighExerciseAmount ∧ Stressed → ReduceTraining |
| HealthyDiet | Overweight ∧ LowExerciseAmount → Healthy Diet ∧ MoreTraining |

Figure 5 depicts the MyData core operations for establishing trust between the components that together realize the recommendation service. The process is the following: (a) An individual links Semantic Reasoner, a health application, and the aggregator service (W2E) to his/her MyData Operator account; (b) The individual authorizes the Semantic Reasoner to access his/her personal health data from the W2E aggregator service; (c) The individual authorizes the linked health application to use data from Semantic Reasoner for health related guidance.

The health application and Semantic Reasoner both have consent tokens for establishing the trust required for data exchange. In this scenario, Semantic Reasoner separates user applications from the aggregated personal data and helps to preserve the original data from being exploited by third party applications.

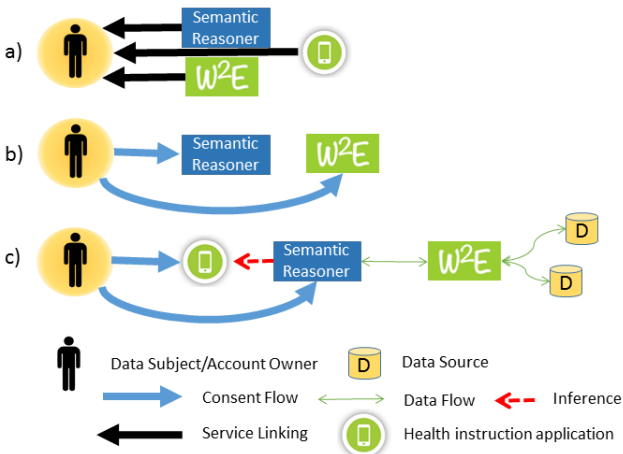

Fig. 5. MyData operations for establishing the process of inferring health related conditions from the wellness data.

## VI. DISCUSSION

This paper suggests providing privacy as a service within distributed, horizontal actor and service environment. The PRIAAS approach facilitates and expedites the creation of new services, data-bindings between actors and endorses new business models to reuse-refine-reuse personal data within individual domain as well as group domain to create new services and application within digital healthcare and wellbeing. For example, the personal data can be seamlessly leveraged among doctors, emergency departments, and family members, to produce efficient and consistent care service, tailored for the individual in question. The proposed PRIAAS architecture also overcomes many expensive system-to-system developments, interoperability and integration problems, introducing consent provisioning between individual and actors in a secure way. Such enablers are needed in making the personal data utilization rights a controlled, but a ubiquitous service, e.g. in prescription, fitness and diet management; and for communication between people having similar health problems, as well as discovering clinical trials to participate in [9]. Advancements on Big Data introduce even more possibilities, on acquiring insight into healthcare related phenomena by analyzing large amounts of unstructured data from large amounts of people, for example previously undervalued electronic health records over aggregated patient collections [10]. Finally, a considerable amount of savings can be expected. As an example, it has been estimated that ability to provide complete interoperability within US health information systems results in savings of $77.8 billion a year [11]; in this, the consent management and privacy is a large factor to optimize costs, privacy and security.

Our PRIAAS architecture offers a holistic solution for consent delivery and management in MyData infrastructure, via clearly defined actors, APIs, and management of consent flows. MyData approach provides benefits for individuals, companies, and the society by allowing data to be used efficiently while still maintain a strong privacy and control within the hands of the individual. MyData addresses the concept of data control rather than data ownership. Individuals gain tools to manage their data and are provided with new innovative services. Companies benefit from the new data-based business opportunities and standardization enables interoperability and lowers the barrier for new companies and businesses to enter. Finally, the society benefits from the standardized structures, processes, and policies addressing the individual's right to their data, as well as the new innovative services.

By giving individuals the power to determine how their data can be used, MyData approach and our architecture enable the collection and use of personal data in ways that maximize the benefits gained and minimize the loss of privacy. GDPR and human-centric principles are complementary to each other. While GDPR aims at strengthening and clarifying practices for data security, the human-centric principles of MyData aim at enabling new services by supporting the use of personal data and providing a controlled and trusted flow from multiple data sources to applications and services. They together provide improved privacy, trust and rights for individuals, as well as aim at unifying protective practices and processes. GDPR will be unifying EU regulations for data protection and MyData will provide GDPR compliant architecture, tools, and practices. GDPR emphasizes unambiguous and informed, for sensitive data also explicit, consent, data usage transparency as well as data portability (interoperability).

Our PRIAAS consent management architecture addresses these requirements with a consent centric framework that enables transparent, human-centered organization of personal data and provides clearly defined interfaces. Semantic technologies are utilized to enrich data and facilitate service interoperability. Moreover, we present a proof-of-concept demonstrating the usage of our privacy-preserving architecture. Our architecture focuses on consents because, first, consents are the primary (but not the only) legislative framework that defines information processing from the human-centric perspective; and, second, human and machine-readable standardized consents unite the technical data management systems, legislative frameworks and the human perspective. Moreover, the same consent management framework can also be used with minor modifications over personal data usage for other fields in addition to health. This research not only increases the level of data protection in general but also suggests a consistent approach for all organizations collecting and processing their data.

The proposed consent management architecture is targeted to the upcoming regulation change. When compared to alternative UMA-based solutions, such as ForgeRock Identity Platform, our novelty is the inclusion of MyData Operator role, which enables trusted transfer of data and consent information with fewer messages by brokering of Sources and Sinks to each other. Therefore, we simplify the authorization flow process with respect to other full UMA-based solutions and hence enable trusted transfer of data and consent information with fewer messages.

The proof-of-concept focuses on managing data and privacy but does not address the other key requirement for human-centric processing and managing of personal information at a sufficient level, hence more work is needed. Our next tasks will be to develop global interoperability and transferability of MyData Accounts and consents based on MyData Operator portability. This requires further standardization and design on trust networks and semantic interoperability. We are targeting also an open business environment, which requires the development and adoption of our architecture and common standards for future MyData operator businesses.

ACKNOWLEDGMENT

This research has been supported by a grant from Tekes as part of Digital Health Revolution programme. The multi-disciplinary programme is coordinated and managed by Center for Health and Technology, University of Oulu, Finland.

REFERENCES

[1] Protection of personal data, European Commission, http://ec.europa.eu/justice/data-protection/
[2] J. Kaye, E.A. Whitley, D. Lund, M. Morrison, H, Teare, and Melham, K. "Dynamic consent: a patient interface for twenty-first century research networks". European Journal of Human Genetics, vol. 23, no. 2, pp.141-146, 2015.
[3] A. Poikola, K. Kuikkaniemi, and H. Honko, "MyData—A Nordic Model for human-centered personal data management and processing" http://urn.fi/URN:ISBN:978-952-243-455-5
[4] The ForgeRock Identity Platform, ForgeRock, https://www.forgerock.com/platform/
[5] T. Hardjono, E. Maler, M. Machulak, and D. Catalano, Kantara Initiative, User-Managed Access (UMA) Profile of OAuth 2.0, https://docs.kantarainitiative.org/uma/rec-uma-core.html.
[6] Minimum Viable Consent Receipt Specification - v 0.7, Kantara Initiative, https://github.com/KI-CISWG/MVCR
[7] R. Arnold, A. Hillebrand, and M. Waldburger, "Personal Data and Privacy", 2015, Ofcom. London.
[8] W2E, https://w2e.fi/frontpage/.
[9] R. Steinbrook, "Personally controlled online health data - the next big thing in medical care?" New England Journal of Medicine, vol. 358, no. 16, pp. 1653-1656, 2008.
[10] T.B. Murdoch, and A.S. Detsky, "The inevitable application of big data to health care". Jama, vol 309, no. 13, pp. 1351-1352, 2013.
[11] J. Walker, E. Pan, D. Johnston, D., J. Adler-Milstein, D.W. Bates, and B. Middleton, "The value of health care information exchange and interoperability". Health Affairs, 24, 2005. W5.